\documentclass[aps,prd,preprint,
superscriptaddress,preprintnumbers,eqsecnum,
showpacs,nofootinbib,nobibnotes]{revtex4}
\usepackage{amsfonts,amssymb,amsmath,bm}
\usepackage{graphicx}

\newcommand{\be}{\begin{equation}}
\newcommand{\bea}{\begin{eqnarray}}
\newcommand{\ee}{\end{equation}}
\newcommand{\eea}{\end{eqnarray}}
\newcommand{\bpi}{\begin{picture}}
\newcommand{\bce}{\begin{center}}
\newcommand{\epi}{\end{picture}}
\newcommand{\ece}{\end{center}}

\def\s#1{{\scriptscriptstyle #1}}
\def\chic#1{{\scriptscriptstyle #1}}
\newcommand{\apt}{{\alpha}_{\mathrm{\chic{PT}}}(q^2)}
\newcommand{\aptz}{{\alpha}_{\mathrm{\chic{PT}}}(0)}

\newcommand{\agh}{\alpha_{\mathrm{\chic{gh}}}(q^2)}
\newcommand{\aptl}{\alpha(q^2)}

\newcommand{\dpt}{\widehat{d}(q^2)}

\newcommand{\Gxi}{G(q^2)}

\newcommand{\Gximu}{G(q^2)}

\def\gb{\bm{\Gamma}}
\def\xiQ{\xi_{\chic{\mathrm Q}}}

\begin{document}

\title{QCD effective charges from lattice data}
%\date{\today}

\author{A.~C. Aguilar}
\affiliation{Federal University of ABC, CCNH, 
Rua Santa Ad\'elia 166,  CEP 09210-170, Santo Andr\'e, Brazil.}
\author{D.~Binosi}
\affiliation{European Centre for Theoretical Studies in Nuclear
  Physics and Related Areas (ECT*), Villa Tambosi, Strada delle
  Tabarelle 286, I-38123 Villazzano (TN), Italy.}
\author{J. Papavassiliou}
\affiliation{Department of Theoretical Physics and IFIC, 
University of Valencia-CSIC,
E-46100, Valencia, Spain.}

\begin{abstract}

We use recent lattice data on the gluon and ghost propagators, as well as  
the Kugo-Ojima function,
in order to extract the non-perturbative 
behavior of two particular definitions of the QCD effective charge, 
one based on the pinch technique construction, and one 
obtained from the standard ghost-gluon vertex. 
The construction relies crucially 
on the definition of two dimensionful quantities, which are invariant under 
the renormalization group, and are built out of 
very particular combinations of the aforementioned Green's functions. 
The main non-perturbative feature of both effective charges, 
encoded in the infrared finiteness of the gluon propagator 
and ghost dressing function used in their definition, 
is the freezing at a common finite (non-vanishing) value, 
in agreement with a 
plethora of theoretical and phenomenological expectations. 
We discuss the sizable discrepancy between the freezing values obtained 
from the present lattice analysis and the corresponding estimates derived 
from several phenomenological studies, 
and attribute its origin to the 
difference in the gauges employed. A particular toy calculation 
suggests that the modifications induced to the non-perturbative 
gluon propagator by the gauge choice may indeed account for the 
observed deviation of the freezing values.

\end{abstract}


\maketitle

\section{Introduction}

In recent years, a large number of independent lattice simulations have furnished 
highly non-trivial information on 
the infrared (IR) behavior of two fundamental ingredients of pure Yang-Mills theories, 
namely the (quenched) gluon and ghost propagators, for both $SU(2)$ and $SU(3)$
~\cite{Cucchieri:2007md,Cucchieri:2009zt,Bogolubsky:2007ud,Bogolubsky:2009dc,Oliveira:2009eh}. 
In particular, these simulations have firmly established that (in the Landau gauge) 
the QCD gluon propagator and the ghost dressing function are IR finite and non-vanishing. 
Given that the entire issue  
is under intense scrutiny, it is natural to  
explore some of the most salient theoretical and phenomenological 
implications of these lattice results.  The purpose of the 
present work is to use the available lattice data 
to extract the running of the 
QCD effective  charge for a wide range of physical momenta, and, in particular,
 its behavior and value in the deep IR. 
This quantity lies   
at the interface between perturbative  and non-perturbative effects in QCD, 
providing a continuous 
interpolation between two physically distinct regimes: the deep ultraviolet (UV), where 
perturbation theory is reliable, and the deep IR, where 
non-perturbative techniques must be employed. 

The generalization  of the  concept of  the renormalization group (RG) invariant  
and process independent effective charge  from QED to
QCD  is far from obvious, and has been discussed extensively in the literature. 
In this article we will consider two 
of the most 
standard definitions of the QCD effective charge. 
The  first charge, to be denoted by $\aptl$, constitutes the 
most direct non-Abelian generalization of the QED effective charge. 
This charge is obtained within the framework of the 
pinch technique (PT)~\cite{Cornwall:1982zr,Cornwall:1989gv,Watson:1996fg},  
and its generalization, known as generalized PT (GPT), introduced in~\cite{Pilaftsis:1996fh}. 
Of particular importance in this construction is the profound correspondence~\cite{Denner:1994nn,Binosi:2002ft} that exists between the PT (GPT)   
and the background-field method (BFM)~\cite{Abbott:1980hw}. 
The second charge, to be denoted  by $\alpha_{\mathrm{gh}}(q^2)$, 
involves the ghost and gluon self-energies, in the Landau gauge, and in the 
kinematic configuration where the well-known Taylor 
non-renormalization theorem \cite{Taylor:1971ff,Marciano:1977su} becomes applicable \cite{Alkofer:2004it}.  

Both effective charges mentioned above display a
strong dependence on the detailed characteristics of some of the most fundamental 
Green's functions of QCD. Specifically, 
in the case of $\agh$ the required ingredients  
are the conventional gluon propagator, $\Delta(q^2)$, (that of the $R_\xi$ gauges) and the ghost dressing function, $F(q^2)$; 
both quantities are simulated on the lattice, and we will use them as inputs for 
obtaining $\agh$. 
For $\aptl$ the situation is slightly more involved.
The fundamental ingredient one needs for obtaining  $\aptl$ 
is the gluon propagator of the PT-BFM, denoted by $\widehat{\Delta}(q^2)$, which, unfortunately, has not been 
simulated on the lattice yet.  
The way one establishes the required connection between the 
conventional gluon propagator (simulated on the lattice) 
and the PT-BFM propagator 
entering into the definition of $\aptl$ is by  
resorting to two powerful non-perturbative identities. First, 
a formal relation known as ``background-quantum'' identity~\cite{Grassi:1999tp, Binosi:2002ez}, given in 
Eq.~(\ref{bqi2}), relates the two gluon propagators by means of a special function, $\Gxi$, which 
plays a central role in the new Schwinger-Dyson equations (SDE)
derived within the PT framework~\cite{Binosi:2007pi}. 
In fact, interestingly enough, in the Landau gauge only, $G(q^2)$ coincides 
with the well-known Kugo-Ojima function~\cite{Kugo:1979gm}. 
The second identity, given in Eq.~(\ref{BRSTcr}), allows one to obtain $\Gxi$ from $F(q^2)$, to a very good approximation, 
given that the function that controls their difference, $L(q^2)$, is numerically rather small, 
and vanishes exactly at $q^2=0$. 
Therefore, even though the theoretical origin of these two effective charges 
is vastly different [{\it e.g.}, $\aptl$ originates from a propagator, while 
$\agh$ from a vertex], they are very close in the entire range of physical momenta,  
and exactly coincide  in the deep IR~\cite{Aguilar:2009nf}.

A large number of theoretical and phenomenological studies, 
based   on  a-priori  very  distinct  approaches~\cite{Cornwall:1989gv,Mattingly:1993ej,Dokshitzer:1995qm,vonSmekal:1997is,Badalian:1999fq,
Aguilar:2002tc,Brodsky:2002nb,Baldicchi:2002qm,Grunberg:1982fw,Gies:2002af,Shirkov:1997wi,
Gracey:2006dr,Prosperi:2006hx} support the  notion of the ``freezing''
of the QCD running coupling in the deep IR. 
In fact, when the QCD charge is constant (non-vanishing!) in the IR, 
and the quark masses are ignored, QCD becomes conformally invariant. 
Therefore, as has been emphasized amply  
in the recent literature~\cite{Brodsky:2003px}, the IR finiteness of the QCD  
effective charge constitutes a crucial requirement for the applicability 
of the powerful AdS/CFT correspondence \cite{Brodsky:2010ur}.   

As  has been argued in numerous works, 
the IR finiteness of the effective charge and that of the  gluon propagator 
are inextricably connected: they can be both traced back to the 
same phenomenon, namely  the non-perturbative generation of a dynamical gluon mass,
through the implementation of the Schwinger mechanism at the level  
of the SDE 
governing the gluon propagator \cite{Aguilar:2008xm}. Within the PT-BFM framework, the SDE solutions for 
the gluon self-energy, denoted by $\widehat\Delta(q^2)$, are used to form 
the RG-invariant combination  $\widehat{d}(q^2)= g^2 \widehat\Delta(q^2)$
which, in turn, may be cast in the form \mbox{$\widehat{d}^{-1}(q^2)  =  
[q^2  +  m^2(q^2)]\{b \ln(\frac{q^2+4 m^2(q^2)}{\Lambda^2})\}$}, 
where $b$ is the first coefficient of the QCD $\beta$ function, and 
$\Lambda$ the QCD mass scale of a few hundred MeV. 
The  non-perturbative  generalization  of  $\alpha(q^2)$,  
the  QCD  effective charge, is contained in 
the curly brackets; evidently, the $m^2(q^2)$ in the argument of the logarithm 
tames  the   Landau pole, and $\alpha(q^2)$ freezes 
at a  finite value in the IR, namely \mbox{$\alpha^{-1}(0)= b \ln (4m^2(0)/\Lambda^2)$}.

%%%%%%%%%%%%%%%%%%%%%%%%%%%%%%%%%%%%%%%%%%%%%%%%%%%%%%%%%%%%%%%%%%%%%%%%%%%%%%%%%%%%%%
The IR finiteness of the effective charge obtained from the lattice data 
becomes manifest in the following way. 
First, one uses the available data for the gluon, the ghost, and the Kugo-Ojima function,
to construct the lattice version of 
the corresponding dimensionful (mass dimension $-2$ ) RG-invariant quantity, 
denoted by $\widehat{d}(q^2)$
in the case of $\aptl$ (as above), and $\widehat{r}(q^2)$
in the case of $\agh$ [see Eq.~(\ref{rgi_pt}) and Eq.~(\ref{gh-RGI}), respectively]. 
The next step is to extract from $\widehat{d}(q^2)$ and $\widehat{r}(q^2)$
a {\it dimensionless} quantity, which will correspond to the associated  
effective charge. 
Both RG-invariant quantities have the 
the gluon propagator, $\Delta(q^2)$, as a common ingredient. Given that  
$\Delta(q^2)$ is effectively massive in the IR, one should follow the standard 
procedure used for massive gauge bosons, such as the $W$ and the $Z$, namely factor out 
a massive ``tree-level'' propagator of the form  $[q^2 + m^2(q^2)]^{-1}$.  
The procedure outlined above guarantees the freezing of the resulting coupling at a finite 
(non-vanishing) value. If, instead, a  $q^{-2}$ is factored out of the 
IR finite gluon propagator, one obtains (trivially) 
an effective charge that vanishes in the IR as $q^{2}$.

%%%%%%%%%%%%%%%%%%%%%%%%%%%%%%%%%%%%%%%%%%%%%%%%%%%%%%%%%%%%%%%%%%%%%%%%%%%%%%%%%%%%%

The article is organized as follows.  In Section II we briefly
review the definitions of the two effective charges under study, 
and recall the fundamental identities, Eq.~(\ref{bqi2}) and Eq.~(\ref{BRSTcr}),  
which relate their ingredients.  
Section III contains the main results of this work.  
In particular, after reviewing some of the most important 
lattice results on the (Landau gauge) gluon and  ghost propagators, 
we construct  the QCD effective  charges and  determine 
their freezing value in the  deep IR. 
In section IV we discuss 
the sizable discrepancy between 
the freezing values obtained in the previous section and those 
favored by a variety of phenomenological studies.
We argue that the main reason for the observed discrepancy is the 
difference in the gauge used: while 
the $\alpha(0)$ extracted from 
the lattice corresponds to the BFM Landau gauge,  
the phenomenological constraints are almost exclusively obtained 
in the BFM Feynman gauge. We will  then derive an approximate formula that
relates the two, suggesting that the discrepancy may be indeed accounted for  by 
difference in gauge choices.  Finally, in Section V we will discuss our results and present our conclusions.

\section{The two effective charges: definitions and basic concepts}

Before introducing the definitions of the effective charges and some of the 
important concepts related to them, we establish the necessary notation. 
The gluon and ghost propagator will be defined as
\bea
\Delta_{\mu\nu}(q)&=&-i\left[ P_{\mu\nu}(q)\Delta(q^2) +\xi\frac{q_\mu q_\nu}{q^4}\right],
\label{prop_cov}\\
D(q^2)&=& \frac{iF(q^2)}{q^2}\,,
\label{ghostdress}
\eea 
where $\xi$ denotes the gauge-fixing parameter, and 
\mbox{$P_{\mu\nu}(q)= g_{\mu\nu} - q_\mu q_\nu /q^2$}
is the usual transverse projector. 
One has $\Delta^{-1}(q^2) = q^2 + i \Pi(q^2)$, 
with  $\Pi_{\mu\nu}(q)=P_{\mu\nu}(q)\Pi(q^2)$ the gluon self-energy; 
finally $F(q^2)$ is the so called ghost dressing function.

A reasonable definition 
of the QCD effective charge may be obtained from the
ghost-gluon vertex  in the  Landau gauge~\cite{vonSmekal:1997is,Alkofer:2004it}. 
Exploiting  the fact that,  in this
gauge,  the ghost-gluon vertex  does   not  get renormalized,  
one  can  construct  the  RG-invariant product               
\be
\widehat{r}(q^2)=g^2(\mu^2)\Delta(q^2) F^2(q^2).
\label{gh-RGI}
\ee
From this quantity one defines the effective charge as 
\be
\agh=[q^2+m^2(q^2)]\widehat{r}(q^2)\,,
\label{alpha-gh}
\ee 
where $\alpha(\mu^2)=g^2(\mu^2)/4\pi$.
Since $\Delta(q^2)$ and $F(q^2)$ corresponds exactly to the quantities measured directly on the lattice, 
this definition constitutes the most direct way of extracting the non-perturbative QCD charge from the lattice. 
It should be noted, however, that away from the  Landau gauge  
additional information on the form-factor of the ghost-gluon vertex must 
be supplemented, in order to define the RG-invariant quantity analogous to the $\widehat{r}(q^2)$ of (\ref{gh-RGI}). 
This necessity, even though is not a limitation of principle, brings about several ambiguities; for example, the aforementioned vertex form-factor depends on two physical momenta, and a particular choice of the scale must be 
implemented, in order for the effective charge to be a function of a single momentum scale. In other words,   
one cannot obtain a universal definition of the charge, {\it i.e.}, one that does not depend on the 
specific kinematic details of the vertex employed. 

A universal (process-independent) definition of an effective charge for every gauge may be obtained from  
the gluon self-energy in the (covariant) BFM. As is well-known,  
this quantity, to be denoted by $\tilde{\Pi}^{(\xiQ)}(q)$, 
captures the running of the QCD coupling for {\it every} value of the (quantum) gauge-fixing parameter, $\xiQ$. 
In particular, at one loop, we have  
\be
i\tilde{\Pi}^{(\xiQ)}(q) =  q^2 g^2 [b \ln\left(-q^2/\mu^2\right) + C_{\xiQ}],
\label{gfo}
\ee
where $b= 11C_{\rm A}/48 \pi^2$ is the first coefficient of the QCD $\beta$ function ($\beta=-bg^3$)  
in the absence of quarks,  $C_{\rm A}$ is the Casimir eigenvalue of the adjoint representation ($C_{\rm A}=N$ for $SU(N)$),  and
the gauge-dependent constant  $C_{\xiQ}$ is given by (third item in \cite{Denner:1994nn}) 
\be
C_{\xiQ} = \frac{C_{\rm A}}{16\pi^2} \left[\frac{(1-\xiQ)(7+\xiQ)}{4} - \frac{67}{9}\right].
\label{Cxi}
\ee
Note that the value $\xiQ=1$,  {\it i.e.}, the Feynman gauge of the BFM, is very special, because it 
reproduces the (gauge-independent) PT gluon self-energy; in this privileged gauge 
{\it all} unphysical longitudinal terms appearing inside 
an ostensibly gauge-independent quantity (physical on-shell amplitude, Wilson-loop, etc) 
have been discarded. 
 
For asymptotically large momenta one may neglect the constant  $C_{\xiQ}$ 
next to the leading logarithm, and write, in any gauge, (Euclidean momenta) 
\be
\widehat{\Delta}(q^2)=\frac1{q^2 [1+bg^2\ln(q^2/\mu^2)]}.
\ee
It is then easy to establish ({\it e.g.}, by resorting to the QED-like identity 
$Z_{\widehat{A}}^{-1/2}=Z_g$, valid in the BFM to all orders and for every $\xiQ$ ~\cite{Abbott:1980hw})
that the product 
\be
\dpt=g^2(\mu^2)\widehat{\Delta}(q^2),
\label{pt-rgi}
\ee 
is invariant under the renormalization group,  {\it i.e.}, it is an RG-invariant quantity, just as the $\widehat{r}(q^2)$ in (\ref{gh-RGI}).
From $\dpt$ one may extract the QCD effective charge exactly as in (\ref{alpha-gh}), namely 
\be
\aptl=[q^2+m^2(q^2)]\dpt. 
\label{alpha-gPT}
\ee

In order to make contact between $\widehat{\Delta}(q^2)$ 
appearing in the definition of the RG-invariant product $\widehat{d}(q^2)$ and the conventional propagator $\Delta(q^2)$
simulated on the lattice (in the Landau gauge), we employ a formal all-order identity,
which relates them as follows~\cite{Grassi:1999tp,Binosi:2002ez}
\be
\Delta(q^2) = 
\left[1+\Gximu\right]^2 \widehat{\Delta}(q^2).
\label{bqi2}
\ee
In the above formula the two gauge fixing constants, $\xi$ and $\xiQ$, associated with $\Delta(q^2)$ and 
$\widehat{\Delta}(q^2)$, respectively,  must be equal (but otherwise arbitrary); 
in particular, in the Landau gauge, $\xi =\xiQ=0$. 

The function $\Gximu$ appearing in (\ref{bqi2}) 
is the $g_{\mu\nu}$ component of a particular two-point function, denoted by $\Lambda_{\mu\nu}(q)$, defined as 
\bea
\Lambda_{\mu\nu}(q) &=& -ig^2C_A
\int_k H^{(0)}_{\mu\rho}
D(k+q)\Delta^{\rho\sigma}(k)\, H_{\sigma\nu}(k,q)
\nonumber\\
&=& g_{\mu\nu} G(q^2) + \frac{q_{\mu}q_{\nu}}{q^2}L(q^2),
\label{LDec}
\eea
where $\int_k\equiv\mu^{\epsilon} (2\pi)^{-d}\int d^dk$, with $d=4-\epsilon$ the space-time dimension.
The function $H_{\sigma\nu}(k,q)$  appears 
in the all-order Slavnov-Taylor identity satisfied by the three gluon vertex~\cite{Ball:1980ax}, and 
is related to the ghost-gluon vertex ${\gb}_{\mu}(k,q)$ 
through the identity 
\be
q^\nu H_{\mu\nu}(k,q)=-i\gb_{\mu}(k,q) \,.
\label{IDHG}
\ee
At tree level, \mbox{$H_{\mu\nu}^{(0)} = ig_{\mu\nu}$} 
and \mbox{$\gb^{(0)}_{\mu}(k,q)=\Gamma_\mu(k,q)=-q_\mu$}. 
Note that both $G(q^2)$ and $L(q^2)$ depend explicitly on the value of the gauge-fixing parameter $\xi$.

Since the origin of the identity in (\ref{bqi2}) is the BRST symmetry of the theory,  it 
does not get deformed by the renormalization procedure. Thus, one can write (\ref{pt-rgi}) in terms of 
$\Delta(q^2)$ and $\Gximu$ as follows
\be
\dpt=\frac{g^2(\mu^2) \Delta(q^2)}{\left[1+\Gximu\right]^2}. 
\label{rgi_pt}
\ee

It turns out ~\cite{Kugo:1995km,Grassi:2004yq,Kondo:2009ug} that the function $\Gximu$, 
coincides (in the Landau gauge only) with the well-known Kugo-Ojima function~\cite{Kugo:1979gm}, $u(q^2)$ defined as   
\be
\int\!d^4x\ \mathrm{e}^{-iq\cdot(x-y)}\langle T\big[\left({\cal D}_\mu c\right)_x^m\left({\cal D}_\mu \bar c\right)_y^n\big]\rangle=
-\frac{q_\mu q_\nu}{q^2}\delta^{mn}+P_{\mu\nu}(q)\delta^{mn} u(q^2),
\label{id-1}
\ee
where $({\cal D}^\mu \Phi)^m=\partial^\mu \Phi^m+gf^{mnr}A^n_\mu \Phi^r$ is the usual QCD covariant derivative.
The Kugo-Ojima function has been simulated on the lattice 
by means of Monte-Carlo averages of the operator time-ordered product appearing on the left-hand 
side of the defining equation~(\ref{id-1})~\cite{Sternbeck:2006rd}. 
Given that $\Gximu=u(q^2)$, the lattice information on $u(q^2)$ may be used, in principle,   
into (\ref{rgi_pt}), together with the lattice results for the Landau gauge $\Delta(q^2)$.  

In addition, $G(q^2)$ is related to the ghost dressing function $F(q^2)$ and the form-factor $L(q^2)$  
of (\ref{LDec}) through the BRST identity
\be
F^{-1}(q^2)=1+G(q^2)+L(q^2).
\label{BRSTcr}
\ee
This identity, in conjunction with the corresponding dynamical equations~\cite{Aguilar:2009nf}
given in Eq.~(\ref{simple}), allows the indirect determination of  $G(q^2)$ and $L(q^2)$ from the 
lattice data on the ghost dressing $F(q^2)$~\cite{Aguilar:2009pp}. 
Thus, provided that one carries out the renormalization procedure 
in  a way that manifestly preserves~(\ref{BRSTcr}), 
the two effective charges  are related through the equation~\cite{Aguilar:2009nf}
\be
\agh=\left[1+\frac{L(q^2)}{1+G(q^2)}\right]^{-2}\aptl.
\label{coupl-id}
\ee
An important corollary of the dynamical equations of Eq.~(\ref{simple}) is that $L(0)=0$. 
In addition,  under  very general  conditions, $G(0)\in(-1,0)$. 
Therefore, from (\ref{coupl-id}) one concludes that~\cite{Aguilar:2009nf}
\be
\alpha_{\mathrm{\chic{gh}}}(0) = \alpha(0).
\label{eqcoupl}
\ee
Since, finally, $\agh$ and $\aptl$ coincide in the deep UV, where
they both reproduce the correct perturbative  behavior,  
the two  charges can
only differ appreciably in the intermediate region of momenta; however, since
$L(q^2)$  is   numerically  suppressed~\cite{Aguilar:2009nf},  this
difference is rather small.

\section{Effective charges from lattice}

This section contains the main results of this article, and is composed of several subsections.
After presenting a collection of lattice data, which firmly establish the IR 
finiteness of the conventional gluon propagator $\Delta(q^2)$ (in the Landau gauge) 
and the ghost-dressing function, we embark on the actual extraction of the effective charges from the 
lattice data, using the definitions and results of the previous section. 
The final results of all the analysis, carried out throughout this section, are shown on the right panel of 
Fig.~\ref{fig7}; evidently,  in the deep IR, both charges, $\agh$ and $\aptl$, saturate at the same  finite value,
as predicted on general principles.  

%%%%%%%%%%%%%%%%%%%%%%%%%%%%%%%%%%%%%%%
%   figure 1
%%%%%%%%%%%%%%%%%%%%%%%%%%%%%%%%%%%%%%%
\begin{figure}[!t]
\begin{center}
\includegraphics[scale=0.8]{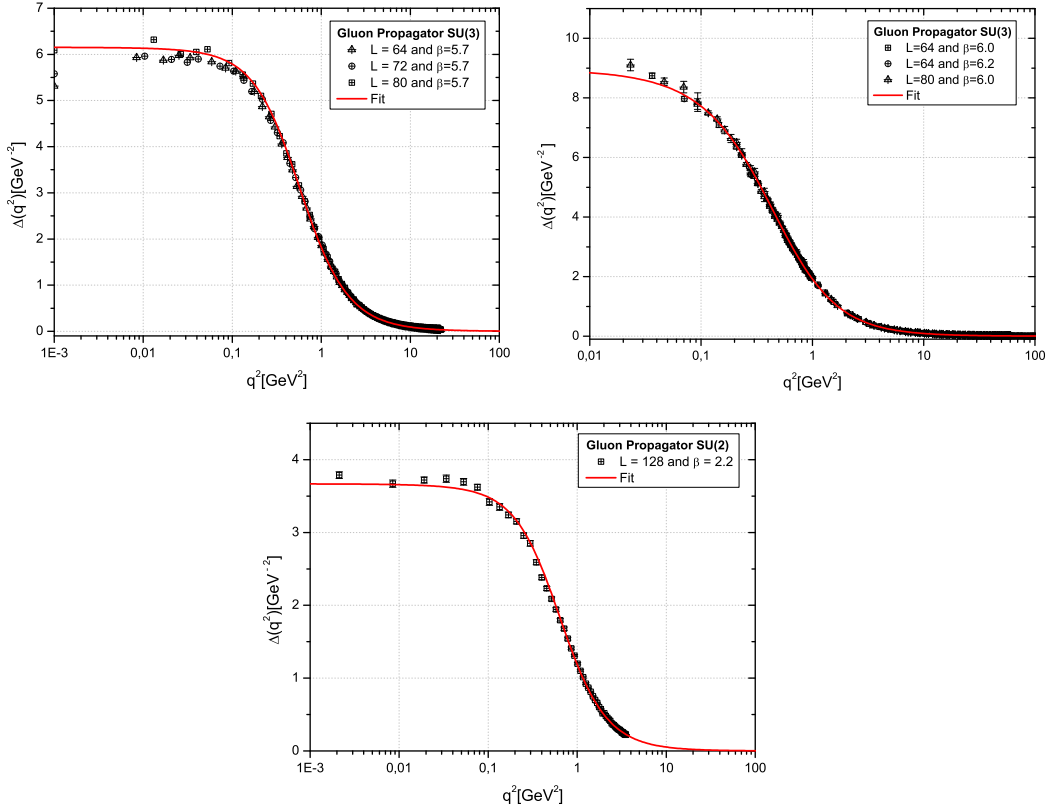}
\end{center}
\vspace{-0.5cm}
\caption{The lattice data for the gluon propagator obtained 
by three independent groups \cite{Cucchieri:2007md, Bogolubsky:2007ud, Oliveira:2009eh} fitted by Eq.~(\ref{fgluon}).  {\it Upper left panel}: Lattice data from Ref.~\cite{Bogolubsky:2007ud} renormalized at \mbox{$\mu=3.0$ GeV}. {\it Upper right panel}: Lattice result,
renormalized at \mbox{$\mu=3.0$ GeV}, obtained in Ref.~\cite{Oliveira:2009eh}. {\it Bottom panel}:  The $SU(2)$ gluon propagator obtained in Ref.~\cite{Cucchieri:2007md}.}
\label{fig1}
\end{figure}

\subsection{Lattice results for the gluon propagator}

In this subsection we present  
some of the most relevant lattice results on the (Landau gauge) gluon propagator, given that 
it constitutes a central common ingredient of both effective charges. 
Even though 
in our analysis we will use only one set of lattice data (that of \cite{Bogolubsky:2007ud}), 
it is important to establish that various groups coincide on the qualitative behavior for the 
Green's functions in question. 
In Fig.~\ref{fig1} we
show the results for the gluon propagator obtained by  
three independent lattice groups \cite{Cucchieri:2007md, Bogolubsky:2007ud, Oliveira:2009eh}. Although, for each group, 
the lattice spacing and the gauge group employed are different, all results 
have as a common feature the appearance of a plateau in the deep IR region, namely 
one of the most salient and distinctive predictions of the 
the gluon mass generation mechanism.
In fact, the three set of data can be accurately fitted in terms of a  massive gluon propagator  of the type
\be
\Delta^{-1}(q^2)= m^2 + q^2\left[1+ \frac{13C_{\rm A}g^2_{f}}{96\pi^2}
\ln\left(\frac{q^2 +\rho\,m^2}{\mu^2}\right)\right]\,,
\label{fgluon}
\ee
where $m^2$, $g^2_{f}$, and $\rho$ are treated as free fitting parameters. 
For the $SU(3)$ lattice simulations, $\mu$ will be chosen to coincide with the 
renormalization point, 
while for the $SU(2)$ case we will treat it as an adjustable parameter.

Specifically, for the data presented on the upper left panel
\cite{Bogolubsky:2007ud},
we find that \mbox{$m^2= 0.16\,\mbox{GeV}^2$}, \mbox{$g^2_{f}=8.79$},
$\rho=4$, and \mbox{$\mu=3\,\mbox{GeV}$}.
For the upper right panel, we use  
\mbox{$m^2= 0.11\,\mbox{GeV}2$}, \mbox{$g^2_{f}=9.77$}, $\rho=9.6$, and
\mbox{$\mu=3\,\mbox{GeV}$}; while
the $SU(2)$ lattice data of Ref.~\cite{Cucchieri:2007md}
can be accurately adjusted using \mbox{$m^2= 0.27\,\mbox{GeV}^2$},
\mbox{$g^2_{f}=27.68$}, $\rho=4$, and \mbox{$\mu=1.92\,\mbox{GeV}$}.

The parameter $m$ acts as a physical  mass 
scale, whose function is  to regulate the perturbative RG logarithm; so, instead of diverging at the Landau pole, the logarithm saturates at a finite value.
Clearly, for large values of $q^2$, we recover the one-loop expression of the gluon propagator in the Landau gauge. 
Note also that for the purposes of this fit we have treated $m$ as if it were a hard mass, even though 
an important theoretical feature of the 
dynamically generated mass is that it should be function of the momentum, vanishing 
in the deep UV in a way consistent with the operator-product expansion, 
displaying either logarithmic or a power-law running \cite{Lavelle:1991ve} [{\it viz.} Eq.(\ref{plr})].

Even though it is evident from Fig.~\ref{fig1} that the various lattice groups 
appear to be in qualitative agreement with each other, 
for the actual extraction 
the effective charges we will rely on the data of 
Ref.\cite{Bogolubsky:2007ud}, given that this latter group uses $SU(3)$ simulations, and has available data also on the ghost propagator.

%%%%%%%%%%%%%%%%%%%%%%%%

\subsection{The remaining lattice ingredients: numerical fits and $\mu$-dependence}

As mentioned earlier, the different definitions of the QCD effective charges
involve three fundamental Green's functions: the gluon propagator $\Delta(q^2)$, the ghost dressing function $F(q^2)$, 
and the auxiliary function $G(q^2)$. In order to verify explicitly the 
expected $\mu$-independence of $\widehat{r}(q^2)$ and $\widehat{d}(q^2)$,    
we need to have at our disposal 
lattice data for $\Delta(q^2)$, $F(q^2)$, and $G(q^2)$ at different  renormalization points.
To that end, we will exploit the property of multiplicative renormalizability,  which allows one to  
connect a set of points renormalized at $\mu$ with the corresponding set renormalized at $\nu$, 
through the relations 
\be
\Delta(q^2,\mu^2)=\frac{\Delta(q^2,\nu^2)}{\mu^2\Delta(\mu^2,\nu^2)}\,, 
\qquad F(q^2,\mu^2)=\frac{F(q^2,\nu^2)}{F(\mu^2,\nu^2)} \,.  
\label{ren_gl}
\ee
Using into Eq.(\ref{ren_gl})
the fundamental 
identity of Eq.~(\ref{BRSTcr}), whose form must be preserved after renormalization, 
we have that $G(q^2)$ must satisfy
\be
1+G(q^2,\mu^2)+ L(q^2,\mu^2)=\frac{1 + G(q^2,\nu^2) + L(q^2,\nu^2)}{1+G(\mu^2,\nu^2)+L(\mu^2,\nu^2)} \,. 
\label{ren_gg}
\ee
Evidently, the self-consistent renormalization procedure of $G(q^2)$ requires the  knowledge of $L(q^2)$.
However, in Ref. \cite{Sternbeck:2006rd}, the renormalization of $G(q^2)$  was carried out considering $L(q^2)$ 
to vanish for all values of momentum,  {\it i.e.}, setting $L(q^2)=0$.  
To be sure, this approximation will not produce any appreciable difference
in the deep IR region, where it was shown~\cite{Kugo:1995km,Aguilar:2009nf} that indeed $L(0)=0$; 
on the other hand, minor changes in the intermediate and UV regimes are to be expected, 
which, however, will be neglected for the purposes of this subsection. $L(q^2)$ will be eventually 
obtained indirectly, 
by substituting lattice data for $F$ and $\Delta$ into the dynamical equation 
of (\ref{simple}); it turns out that it is indeed numerically suppressed within the entire range of available momenta, see Fig.~\ref{fig6}.   

\begin{figure}[!t]
\begin{center}
\includegraphics[scale=0.8]{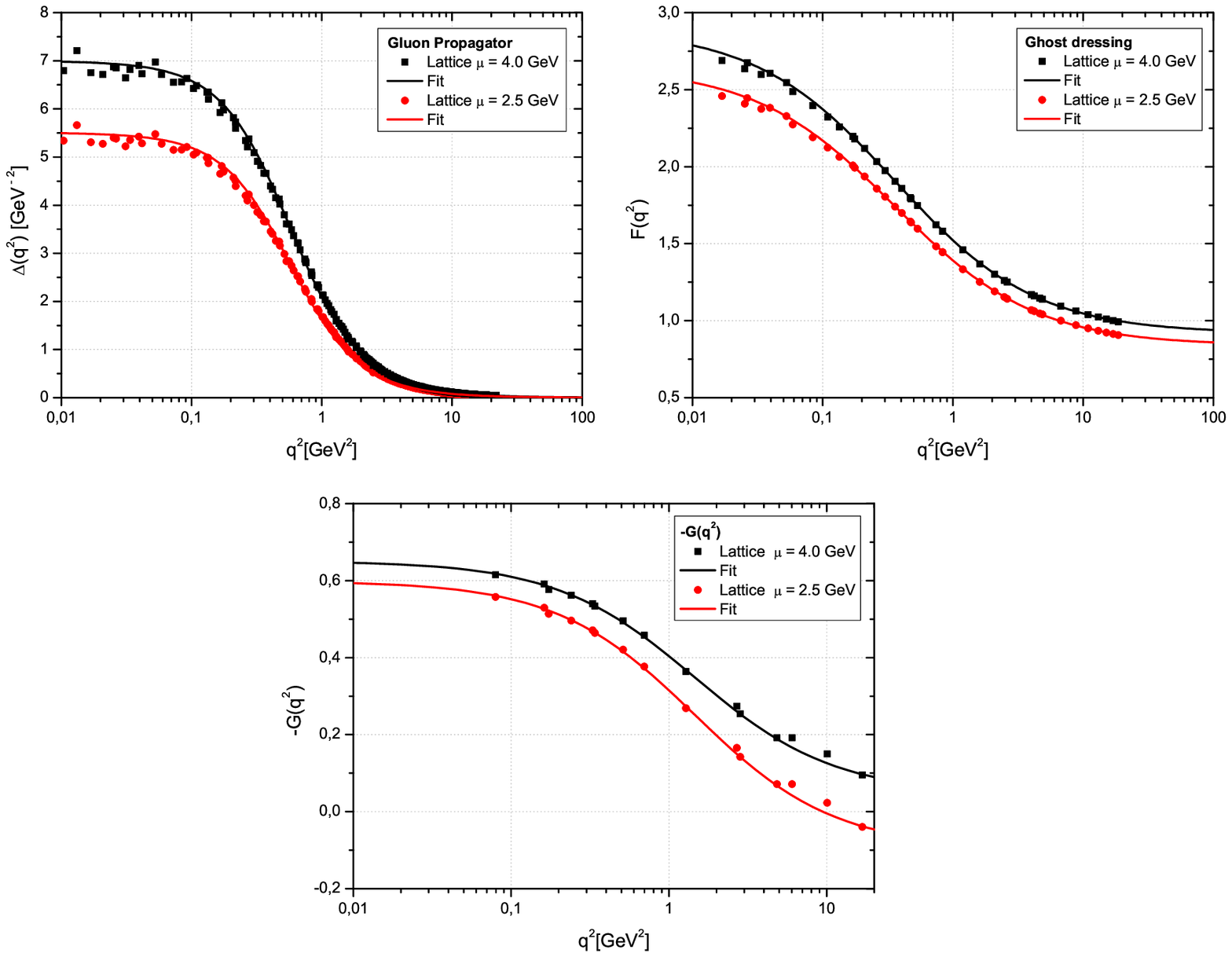}
\end{center}
\vspace{-0.5cm}
\caption{Lattice results for the gluon propagator (upper left panel), the ghost dressing function (upper right panel)
and $-G(q^2)$ (bottom panel) renormalized at  \mbox{$\mu = 2.5 \,\mbox{GeV}$} (red curve) and \mbox{$\mu = 4.0 \,\mbox{GeV}$} (black curve). Values of the fitting parameters of Eqs.~(\ref{fgluon}) and (\ref{fghost}) are: \mbox{$m^2= 0.18\, \mbox{GeV}^2$}, \mbox{$g^2_{f}=10.64$}, \mbox{$a_1=2.64$}, \mbox{$a_2=0.84$}, \mbox{$b_1=0.58$}, and \mbox{$b_2=-0.14$}, for $\mu=2.5$ GeV; \mbox{$m^2=0.14\, \mbox{GeV}^2$}, \mbox{$g^2_{f}=6.95$},  \mbox{$a_1=2.89$}, \mbox{$a_2=0.91$}, \mbox{$b_1=0.65$}, and \mbox{$b_2=0.047$}, for $\mu=4.0$ GeV. For both  values of $\mu$ we use $\rho=4$, \mbox{$p_1=0.8$}, \mbox{$q_1^2=0.36 \, \mbox{GeV}^2$}, \mbox{$p_2=0.98$}, and  \mbox{$q_2^2=1.45 \, \mbox{GeV}^2$.} }
\label{fig2}
\end{figure}

Thus, considering for now $L(q^2)=0$, and  choosing the two 
different values $\mu=2.5$ GeV, and $\mu=4.0$ GeV, we obtain the curves
for $\Delta(q^2)$, $F(q^2)$, and $G(q^2)$  shown in Fig.~\ref{fig2}. 

As can be seen from this figure, the lattice sets used are restricted to momenta ranging roughly from $0.01$ to  \mbox{$22\,\mbox{GeV}^2$}.  Within this range, the gluon propagator can be fitted by Eq.~(\ref{fgluon}), 
while $F(q^2)$ and $G(q^2)$ are being given by  
\begin{equation}
F(q^2)= \frac{a_1-a_2}{1+\left(q^2/q_1^2\right)^{p_1}} + a_2 ,
\qquad 
G(q^2)= \frac{-b_1+b_2}{1+\left(q^2/q_2^2\right)^{p_2}} - b_2,
\label{fghost}
\end{equation}
with the values of the fitting parameters quoted in the caption of Fig.~\ref{fig2}.

%%%%%%%%%%%%%%%%%%%%%%%%%%%%%%%%%%%
%    figure 2
%%%%%%%%%%%%%%%%%%%%%%%%%%%%%%%%%%%
%%%%%%%%%%%%%%%%%%%%%%%%%%%%%%%%%%%%%%%%%%

Notice that in the case of the ghost dressing function, the lattice data, and correspondingly our 
fit, show no enhancement in the deep IR; instead, $F$
saturates at the constant value $a_1$ (in agreement with the large-volume lattice simulations). In addition, 
the Kugo-Ojima confinement criterion is clearly not satisfied, since  $G(0)$ deviates appreciably 
from the special value of $-1$.

\subsection{Fixing the value of $g^2(\mu)$}

%%%%%%%%%%%%%%%%%%%%%%%%%%%%%%%%
%    Figure 3
%%%%%%%%%%%%%%%%%%%%%%%%%%%%%%%
\begin{figure}[!t]
\begin{minipage}[b]{0.45\linewidth}
\centering
%\hspace{-1cm}
\includegraphics[scale=0.45]{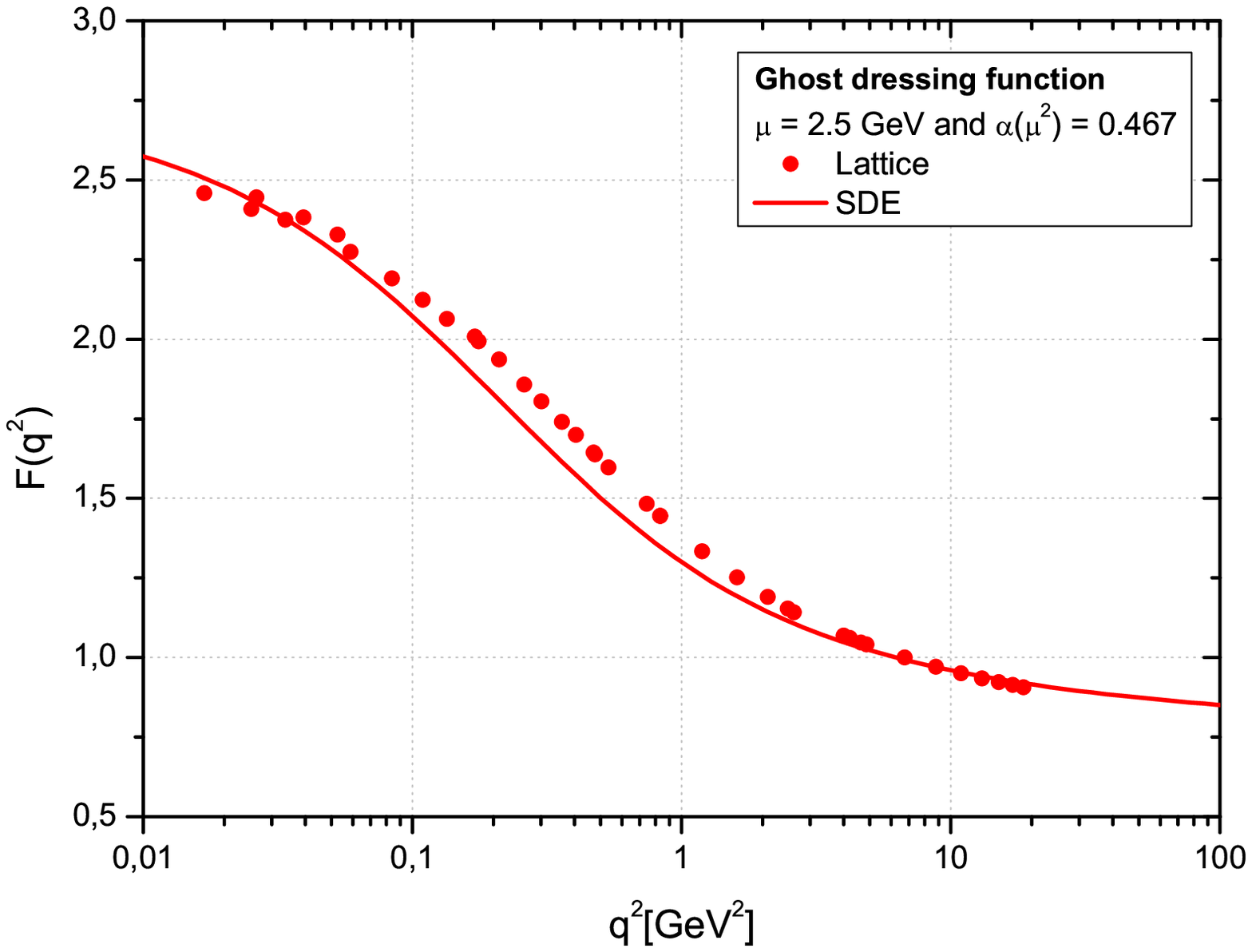}
\end{minipage}
\hspace{0.5cm}
\begin{minipage}[b]{0.50\linewidth}
\centering
%\hspace{-1.5cm}
\includegraphics[scale=0.45]{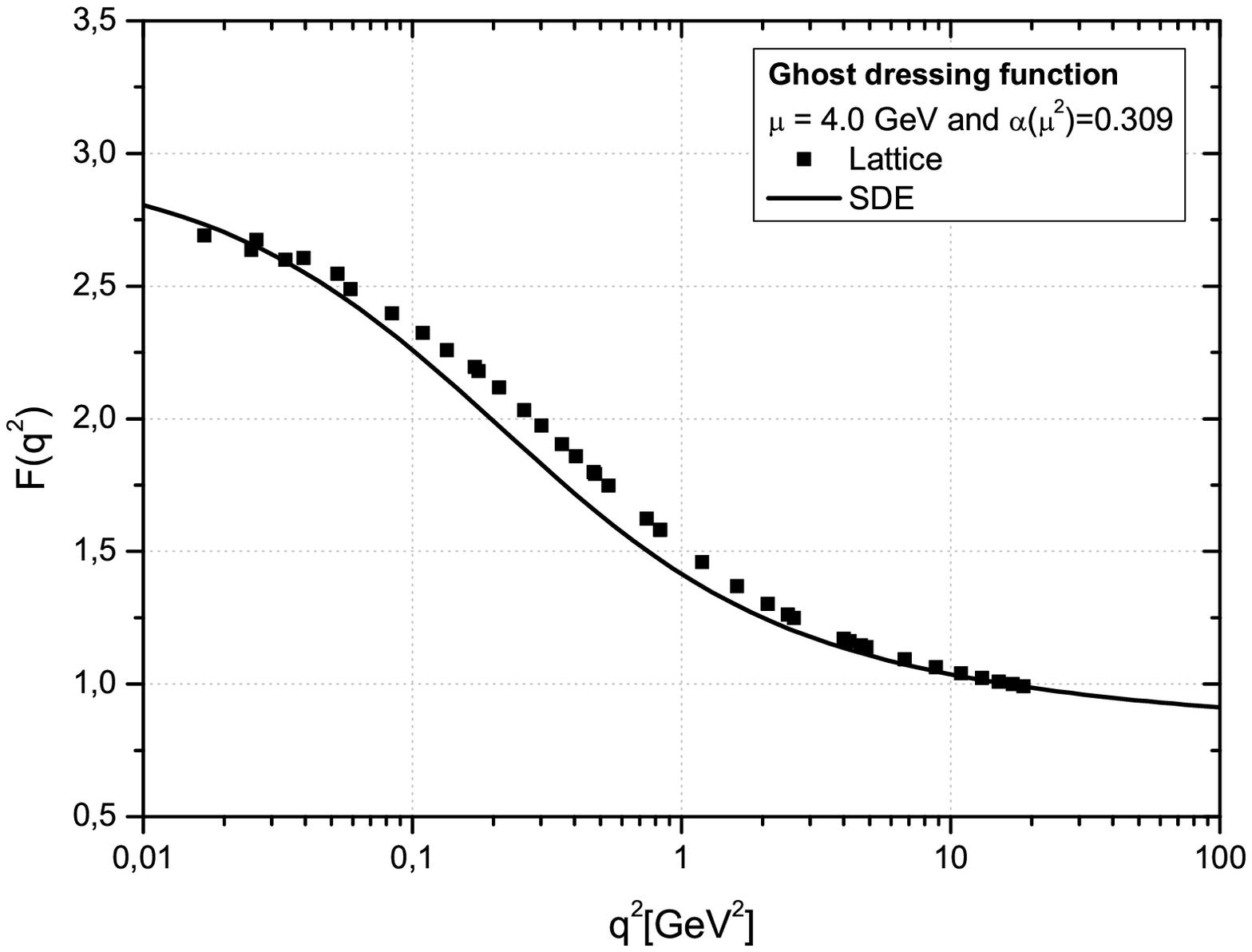}
\end{minipage}
\vspace{-1.0cm}
\caption{Comparison between the ghost dressing function $F(q^2)$ obtained from the ghost SDE (continuous lines) and the corresponding lattice data at  \mbox{$\mu = 2.5 \,\mbox{GeV}$} (left panel) and \mbox{$\mu = 4.0 \,\mbox{GeV}$} (right panel).}
\label{fig3}
\end{figure}

The next step is to determine the value of the renormalized coupling $g(\mu^2)$ 
that enters in both the definitions of the two RG-invariant quantities, $\widehat{r}(q^2)$ and $\widehat{d}(q^2)$.  
To that end, we resort to  
the three (renormalized) dynamical equations for $F(q^2)$, $G(q^2)$, and $L(q^2)$~\cite{Aguilar:2009nf}, namely
\bea
F^{-1}(q^2) &=& Z_c +g^2 C_{\rm {A}} \int_k\left [1-\frac{(k\cdot q)^2}{k^2q^2}\right] \Delta (k)  D(k+q),
\nonumber\\\nonumber \\
1+G(q^2) &=& Z_c + \frac{g^2 C_{\rm {A}}}{d-1}\int_k\left[
(d-2)+ \frac{(k\cdot q)^2}{k^2q^2}\right]\Delta(k)  D(k+q),
\nonumber\\
L(q^2) &=& \frac{g^2 C_{\rm {A}}}{d-1}\int_k 
\left[1 - d \frac{(k\cdot q)^2}{k^2q^2}\right]\Delta(k)  D(k+q).
\label{simple}
\eea
The above equations have been derived using tree-level values for the two fully dressed vertices appearing in them, 
namely the conventional ghost-gluon vertex $\gb_{\mu}$ and the kernel $H_{\mu\nu}$ [{\it viz.} Eq.(\ref{LDec})].  
This appears to be a good approximation,  given that (i) $\gb_{\mu}$
has been studied in lattice simulations~\cite{Cucchieri:2004sq}, 
where it was found to deviate only mildly from its tree-level value, and (ii) $\gb_{\mu}$ and  $H_{\mu\nu}$ are connected 
by the identity~(\ref{IDHG}). 
The renormalization constant  $Z_c$ is determined by the condition 
$F(\mu^2) =1$. %, where $\mu$ is the renormalization point. 
Notice that the renormalization procedure followed~\cite{Aguilar:2009nf} 
preserves the form of the crucial BRST identity Eq.~(\ref{BRSTcr}), as required. 
The most immediate consequence of this renormalization procedure is that the value of $G(\mu^2)\neq 0 $; in fact, 
$G(\mu^2)=-L(\mu^2)$ [see also the discussion following  Eq.~(\ref{ren_gg})]. 

The way the value of $g(\mu^2)$ is determined from Eq.~(\ref{simple}) is the following. 
One substitutes into the integrals on the rhs the lattice data for $\Delta$ and $D$, carries out the 
integration numerically, and adjusts the value of the $g^2$ multiplying the integrals such that 
the result of the integration coincides as well as possible with the available lattice data on $F$ and $G$.

\begin{figure}[!t]
\begin{minipage}[b]{0.45\linewidth}
\centering
%\hspace{-1cm}
\includegraphics[scale=0.45]{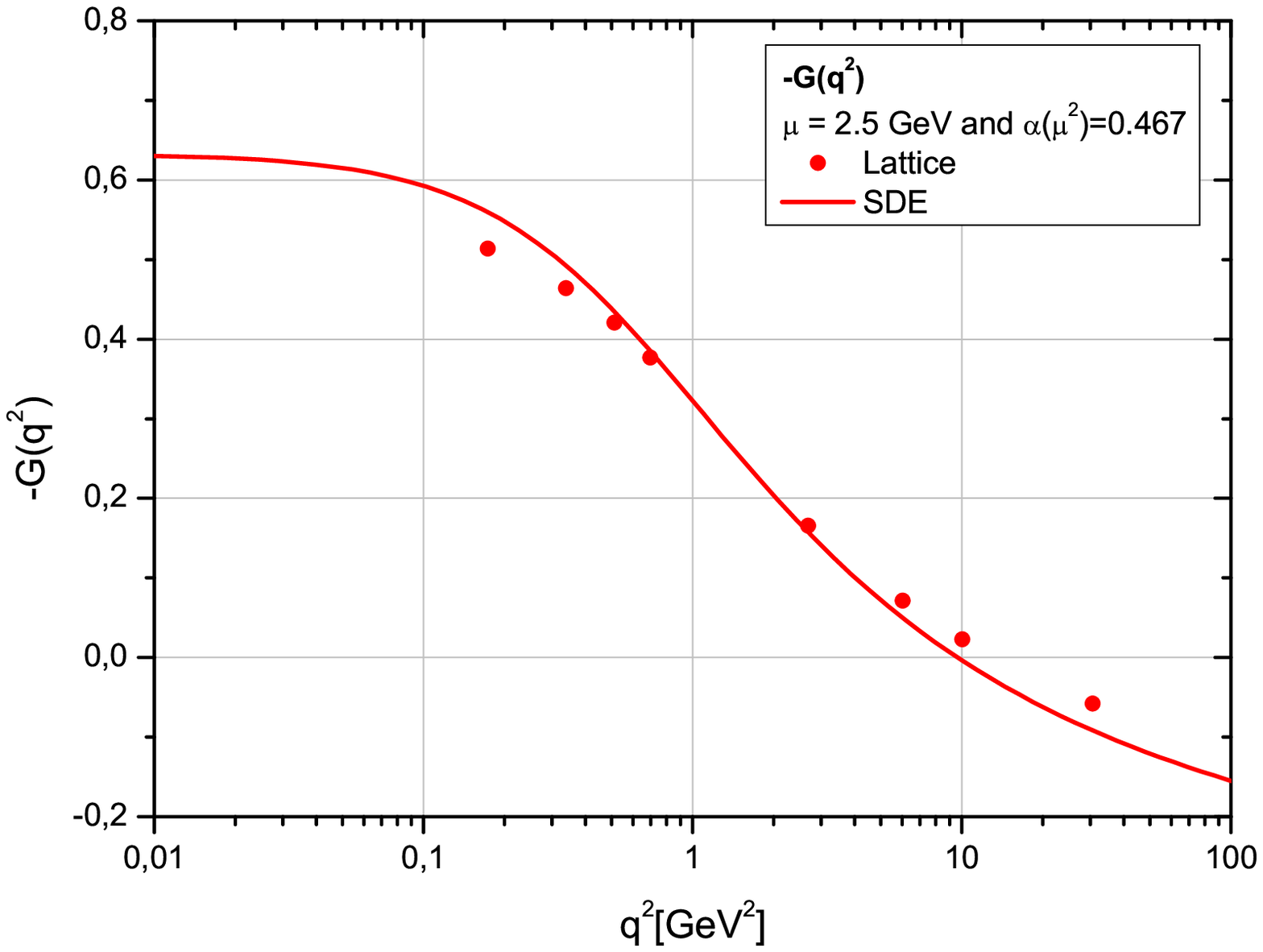}
\end{minipage}
\hspace{0.5cm}
\begin{minipage}[b]{0.50\linewidth}
\centering
%\hspace{-1.5cm}
\includegraphics[scale=0.45]{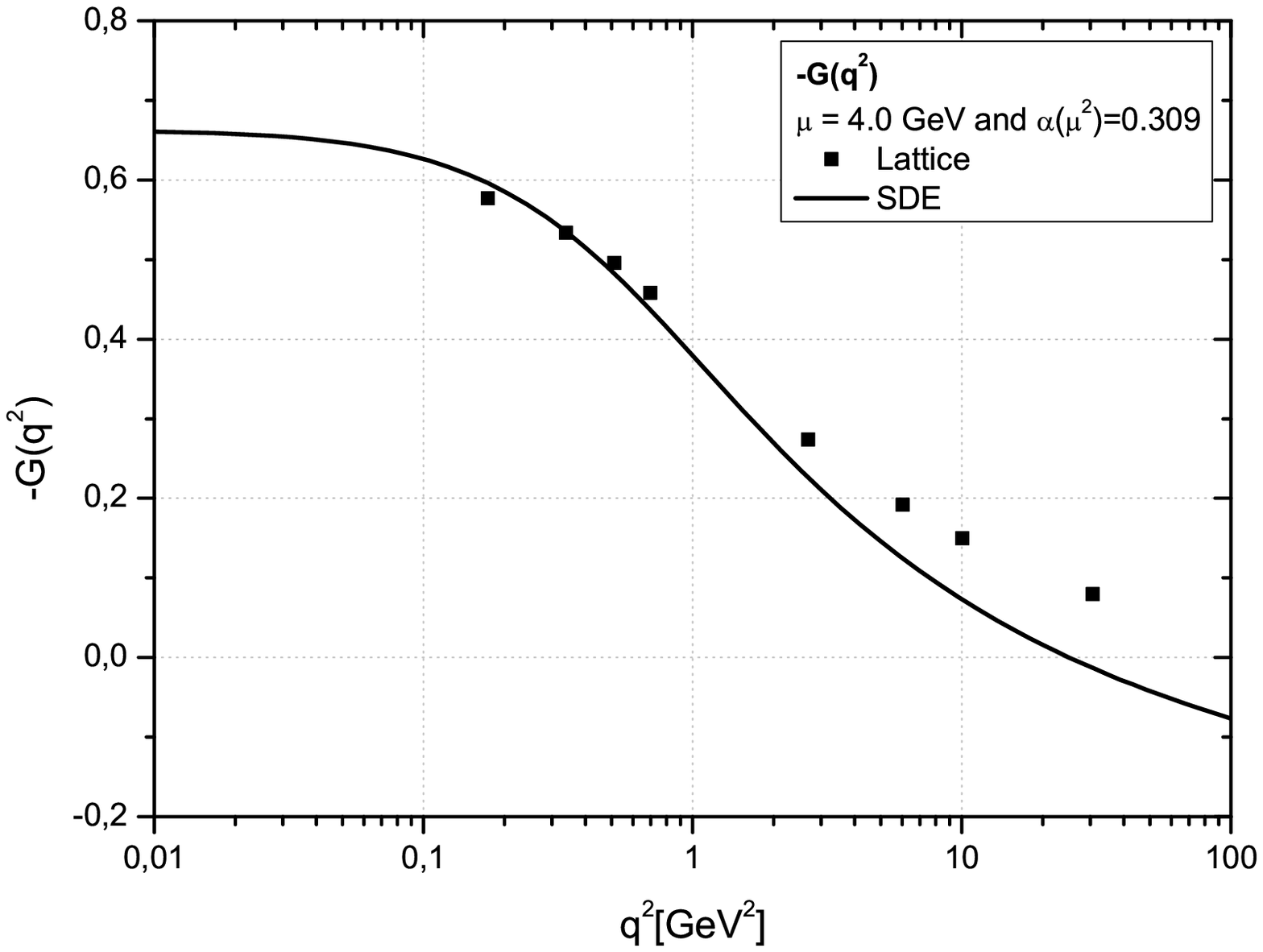}
\end{minipage}
\vspace{-1.0cm}
\caption{Comparison between the $G(q^2)$ function  obtained from the SDE (continuous lines) and the corresponding lattice data at  \mbox{$\mu = 2.5 \,\mbox{GeV}$} (left panel) and \mbox{$\mu = 4.0 \,\mbox{GeV}$} (right panel).}
\label{fig4}
\end{figure}

The results of this procedure for the ghost dressing function $F(q^2)$ are presented in Fig.~\ref{fig3},  
and for $G(q^2)$ in Fig.~\ref{fig4}; in particular, we obtain the value $\alpha(\mu^2)= 0.467$ for $\mu=2.5$ GeV, 
and  $\alpha(\mu^2)= 0.309$ for $\mu=4.0$ GeV. 
%%%%%%%%%%%%%%%%%%%%%%%%%%%%%%%%
%    Figure 4
%%%%%%%%%%%%%%%%%%%%%%%%%%%%%%%

\begin{figure}[t]
\begin{center}
\includegraphics[scale=0.5]{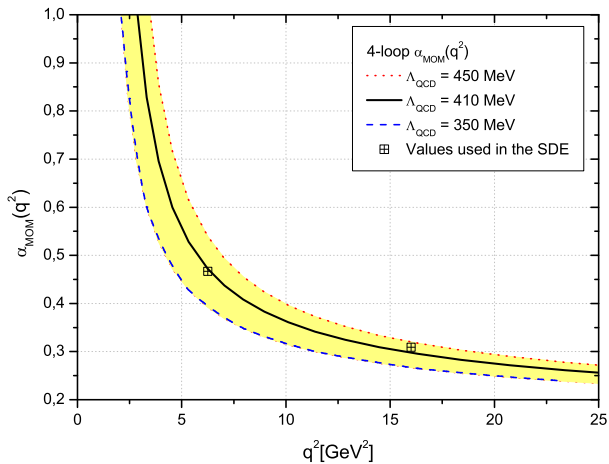}
\end{center}
\vspace{-1.0cm}
\caption{The perturbative running coupling in the MOM scheme, $\alpha_{\rm {MOM}}(q^2)$, up to four-loops for different
values of $\Lambda_{\s{\mathrm{QCD}}}$. The black squares represent the values we use for $\alpha(\mu^2)$.}
\label{fig5}
\end{figure}

In order to check if the values found for $\alpha(\mu^2)$ through the above procedure 
are compatible with what one would expect within the momentum subtraction scheme (MOM) that we use, 
we compare them with the corresponding four-loop perturbative  
calculation presented in~\cite{Boucaud:2005rm}. The result of this comparison is shown  
in Fig.~\ref{fig5}; the yellow band is obtained by varying the \mbox{$\Lambda_{\s{\mathrm{QCD}}}$}, 
appearing in the expression derived in~\cite{Boucaud:2005rm}, in the range between $350-450$ MeV.
As we can see, the best adjustment for the values of $\alpha(\mu^2)$ occurs for \mbox{$\Lambda_{\s{\mathrm{QCD}}}=410\,\mbox{MeV}$}.
%%%%%%%%%%%%%%%%%%%%%%%%%%%%%%%
%    Figure 5
%%%%%%%%%%%%%%%%%%%%%%%%%%%%%%%

\subsection{Final results}

From    all   the   ingredients presented so far, one may construct the 
two RG-invariant   quantities, 
$\widehat{r}(q^2)$ and $\dpt$ of Eqs~(\ref{gh-RGI}) and~(\ref{rgi_pt}). 
A crucial check of the self-consistency of the entire procedure is 
the numerical verification of the theoretically expected independence of the 
above quantities of the renormalization point $\mu$. 
To verify this important point, $\widehat{r}(q^2)$ and $\dpt$ 
have been calculated 
using into the defining equations two different 
sets of inputs for $\Delta$, $F$, and $G$, one  set renormalized at \mbox{$\mu$=4.0 GeV}, and another renormalized at \mbox{$\mu$=2.5 GeV}.
The values for $\alpha(\mu^2)$ are precisely those obtained through the procedure 
of the previous subsection, namely $\alpha(\mu^2)= 0.467$ for \mbox{$\mu=2.5$ GeV}, 
and  $\alpha(\mu^2)= 0.309$ for \mbox{$\mu=4.0$ GeV}. 
The results of this construction are shown 
on the left panel of Fig.~\ref{fig7}; clearly, the $\widehat{r}(q^2)$ and $\dpt$ obtained from each set of data 
are practically on top of each other, thus numerically confirming the theoretical expectations.  
One can also see that the two quantities behave as expected, differing only in the intermediate region of momenta (20 -- 600 MeV).

 %%%%%%%%%%%%%%%%%%%%%%%%%%%%%%%
%    Figure 6
%%%%%%%%%%%%%%%%%%%%%%%%%%%%%%%
\begin{figure}[t]
\begin{center}
\includegraphics[scale=0.5]{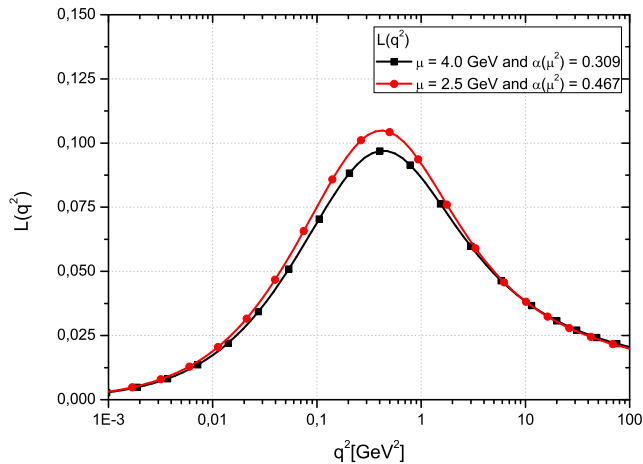}
\end{center}
\vspace{-1.0cm}
\caption{$L(q^2)$ determined from the corresponding SDE~(\ref{simple}), using the solutions 
for $\Delta(q^2)$ and $F(q^2)$ presented in the Fig.~\ref{fig2} respectively, at the same renormalization point. }
\label{fig6}
\end{figure}
Next, using as ingredients the lattice data for  $\Delta(q^2)$ and $F(q^2)$ presented
in Fig.~\ref{fig2} and the values of $\alpha(\mu^2)$ quoted in Fig.~\ref{fig5}, 
one can compute, for the sake of completeness, the auxiliary function $L(q^2)$ from the SDE~(\ref{simple}). The results for $L(q^2)$
are presented in Fig.~\ref{fig6} for both values of $\mu^2$. 
From Fig.~\ref{fig6}, it is easy to check the three properties of $L(q^2)$ mentioned before: (i) indeed $L(q^2)$ is 
numerically rather small over the full range of momenta, (ii) it vanishes in the deep IR, and (iii) its maximum occurs in the intermediate momenta
region (around $500$ MeV).

At this point,  the non-perturbative running charges, $\agh$, and $\aptl$, defined in Eqs.~(\ref{alpha-gh}) and (\ref{alpha-gPT}), 
respectively, 
may be extracted  by multiplying the results obtained for the corresponding RG-invariant quantities by the factor $[q^2 + m^2(q^2)]$. 
To do that, however, one must assume a functional form for the running mass  $m^2(q^2)$. We will use
a mass that decreases in the UV  as power-law running (see, {\it e.g.}, ~\cite{Lavelle:1991ve,Dudal:2008sp})
\be
m^2(q^2)= m^4_0/(q^2+m^2_0)\,.
\label{plr}
\ee
The running mass of (\ref{plr}) 
has a finite value  at $q^2\to 0$, {\it i.e.} $m^2(0)=m^2_0$, with 
a power-law decrease in the deep UV. 
For $m_0$ we choose some representative values  consistent with the phenomenological studies, 
namely  \mbox{$m_0= (500 - 600)$ MeV}  \cite{Cornwall:1982zr,Parisi:1980jy,Cornwall:2009ud}. 

The effective charges obtained following the above steps are shown in the right panel of Fig.~\ref{fig7}.
Evidently, both charges exhibit the  correct (UV) perturbative
behavior, and  freeze at  the same finite  IR values  corresponding to
\mbox{$\alpha_{\mathrm{\chic{gh}}}(0)=\alpha(0)=4.45$} ($m_0=500$  MeV) and 
\mbox{$\alpha_{\mathrm{\chic{gh}}}(0)=\alpha(0)=6.40$} ($m_0=600$  MeV). 
The difference 
between the two couplings is  only in the intermediate momenta region,
and  it is  entirely  due to  the  $L(q^2)$ function;  in this  region
$\aptl$ is always bigger than $\agh$.

%%%%%%%%%%%%%%%%%%%%%%%%%%%%%%%%
%    Figure 7
%%%%%%%%%%%%%%%%%%%%%%%%%%%%%%%
\begin{figure}[!t]
\begin{minipage}[b]{0.45\linewidth}
\centering
%\hspace{-1cm}
\includegraphics[scale=0.45]{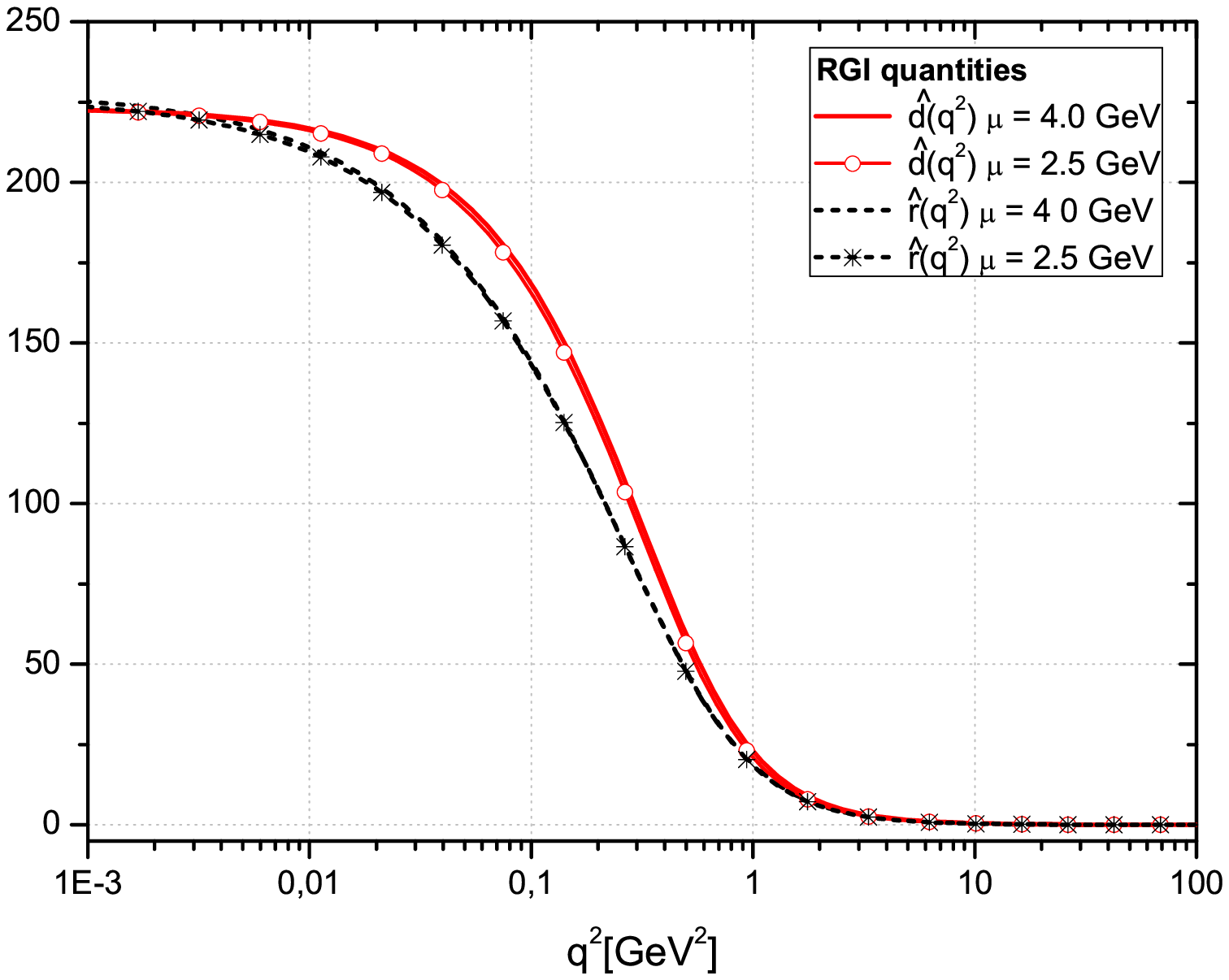}
\end{minipage}
\hspace{0.5cm}
\begin{minipage}[b]{0.5\linewidth}
\centering
\hspace{-0.5cm}
\includegraphics[scale=0.45]{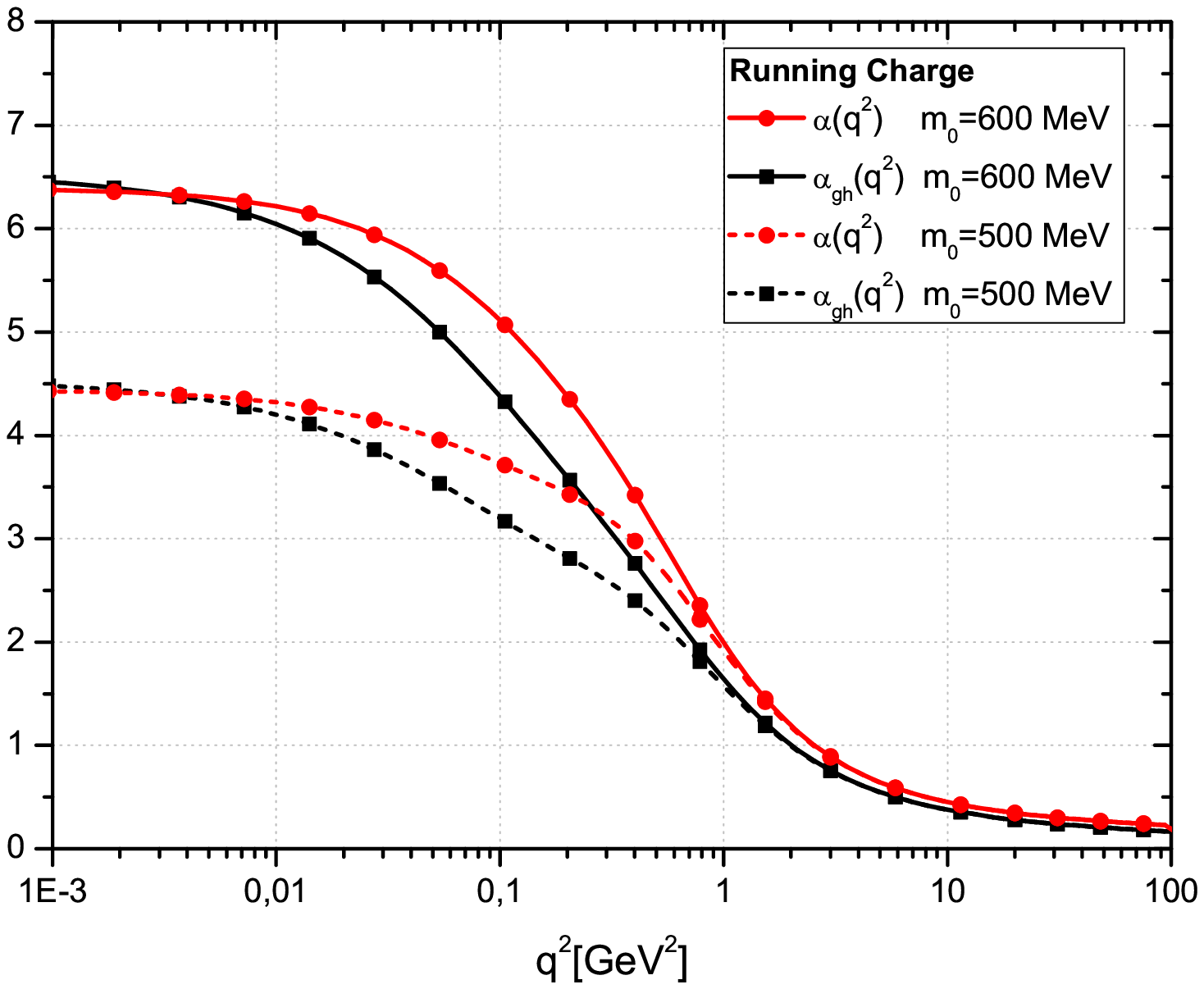}
\end{minipage}
\vspace{-.5cm}   
\caption{\label{fig7}
{\it Left panel}: Comparison between the two RG-invariant products $\dpt$ (solid line) and $\widehat{r}(q^2)$ (dashed line); 
notice that there are two overlapping curves at different $\mu$ for each product. {\it Right panel}: 
Comparison between the QCD effective charge extracted from lattice data:  $\aptl$ (red line with circles) 
and $\agh$ (black line with squares) for two different masses: $m_0=500$ MeV (dashed) and $m_0=600$ MeV (solid).}
\end{figure}

\section{Reconciling lattice with phenomenology}

The effective charges we have obtained from the lattice 
(within the MOM renormalization scheme that we use), reach values in the deep IR that 
are almost an order of magnitude higher than those  
obtained from a large number of phenomenological 
studies. In particular, while the charge obtained from the lattice ranges between 4.5~--~6.5 
[depending on the value of $m(0)$],  
the systematic fitting of numerous processes suggests values for the effective charge in the range 
$0.7\pm0.3$ (for a similar range of gluon masses).

If one were to take both lattice and  phenomenological 
results at face value, one should attempt to determine  
the reason for this sizable discrepancy. 
In this section we will address this issue in the context of a toy calculation,
and we will argue that the observed discrepancy may be traced 
back to the difference in the gauge used when extracting the lattice results 
(the Landau gauge of the BFM, $\xiQ=0$) and that assumed in the 
phenomenological studies (the Feynman gauge of the BFM, $\xiQ=1$).  
Even though we cannot reach firm conclusions, 
our calculation seems to  
indicate that the difference in the gauges may indeed 
reconcile lattice with phenomenology.

The crux of the matter is that the effective charge entering 
into physical processes is neither $\agh$ nor $\aptl$, but rather 
the genuine PT effective charge~\cite{Cornwall:1982zr,Cornwall:1989gv,Watson:1996fg}, to be denoted by $\apt$. 
This charge is defined exactly as $\aptl$ in (\ref{alpha-gPT}), but with the 
crucial difference that the propagator used to form the $\dpt$ 
is the PT gluon propagator,  {\it i.e.}, the BFM propagator calculated 
in the Feynman gauge.  
The  Feynman gauge  of the BFM is privileged,  
in  the  sense that  it  is  selected  {\it dynamically} when  the gluon self-energy  
is embedded into  a physical
observable (such as an on-shell test-amplitude).   Specifically, this gauge 
captures the net propagator-like subamplitude emerging after QED-like
properties have been replicated inside the test-amplitude, by means of
the PT  procedure. Therefore, any gauge-related exchanges 
between the Green's functions put together  to form observables, are eliminated in this particular gauge. 
Instead, the gluon propagator in the Landau gauge, for example, 
contains still residual unphysical contributions, which, when introduced into a physical 
amplitude, will cancel against similar terms from vertex and box diagrams (see third item in \cite{Denner:1994nn}). 

For asymptotically large momenta 
the numerical difference between the charges defined in either gauge
is controlled by the constant $C_{\xiQ}$, given in (\ref{Cxi}).
Evidently, in the UV this difference is subleading, and cannot give rise to any 
appreciable difference. Non-perturbatively, however, the 
difference between the two charges may be sizable. If, for example, we subscribe to the 
notion of dynamical mass generation, a difference in the gauge may lead to a vastly 
different IR behavior. 
In   order   to  gain   a   quantitative
understanding of  how the difference in the  gauge used  
may cause a significant disparity in the infrared values of the corresponding effective charges, 
we consider a model where the gauge bosons are endowed with a mass at tree-level. 
This will allow us to calculate, at  one-loop level, 
the deviation between the two propagators, and the 
discrepancy that it induces  to $\alpha(0)$ and $\alpha_{\mathrm{\chic{PT}}}(0)$. 
The model in question is simply the electroweak sector of the Standard Model, with the 
electric charge set to zero, or, equivalently, with  
$\sin\theta_{\s{\mathrm{W}}} =0$, where $\theta_{\s{\mathrm{W}}}$ is the electroweak mixing (Weinberg) angle.   
In this limit the three gauge bosons (two $W$s and one $Z$) are degenerate. 

At one-loop, the $SU(2)$ gluon self-energy, or equivalently, the $Z$-boson self-energy, 
may be obtained from the results of~\cite{Weiglein:1994yy}, for any value of $\xi_\chic{Q}$.  
Specifically, one has (in Minkowski space)
\bea
\widehat{\Pi}_{\xi_\chic{Q}}(q^2)&=&
{\Pi}_{\mathrm{{F}}}(q^2)+\frac{g^2}{4(4\pi)^2}\frac{q^2-m^2}{(1-d)m^4}H_{\xi_\chic{Q}}(q^2),\nonumber \\
H_{\xi_\chic{Q}}(q^2)&=&\frac{2m^2}{q^2}\left[m^2+\left(9-4d\right)q^2-\left(q^2+m^2\right)\xiQ\right]\left[A_0(\xiQ m^2)-A_0(m^2)\right]\nonumber \\
&-&\left[4\left(5-2d\right)m^4-8\left(2-d\right)m^2q^2+m^2q^2+q^4\right]B_0(q^2;m^2,m^2)\nonumber \\
&+&2\frac{q^2+m^2}{q^2}\left[\left(1-\xiQ\right)^2m^4-2\left(3-2d+\xiQ\right)m^2q^2+q^4\right]B_0(q^2;m^2,\xiQ m^2)\nonumber \\
&-&\left(q^2+5m^2\right)\left(q^2-4\xiQ m^2\right)B_0(q^2;\xiQ m^2,\xiQ m^2),
\eea
where ${\Pi}_{\mathrm{{F}}}(q^2)\equiv {\Pi}_{(\xi=1)}(q^2)$ is the {\it conventional}
gluon self-energy in the Feynman gauge, 
$m$ denotes the effective gauge boson mass, and $A_0$ and $B_0$ are given by 
\bea
A_0 (m^2) &=& 16 \pi^2  \int_k \frac{1}{k^2-m^2} \,,
\nonumber\\
B_0 (q^2;m_1^2,m_2^2)&=& 16 \pi^2 \int_k \frac{1}{(k^2-m_1^2)[(k+q)^2-m_2^2]}\,.
\label{PV}
\eea
Setting $\xiQ=1$ in the above formula we recover the standard 
PT result for the one-loop self-energy of the $Z$-boson~\cite{Degrassi:1992ue} , to be denoted by $\widehat{\Pi}_{\mathrm{{F}}}(q^2)$, namely  
\be
\widehat{\Pi}_{\mathrm{{F}}}(q^2) = {\Pi}_{\mathrm{{F}}}(q^2) - \frac{g^2}{4 \pi^2} (q^2-m^2) B_0(q^2;m^2,m^2)\,.
\label{PTZ}
\ee
Let us now take the difference $R(q^2)$ between 
$\widehat{\Pi}_{\xi_\chic{Q}}(q^2)$ calculated in the Landau and Feynman gauges ($\xi_\chic{Q}=0$, and $\xi_\chic{Q}=1$, respectively); 
denoting the former by $\widehat{\Pi}_{\mathrm{{L}}}(q^2)$, one has in the limit $q^2\to0$ and $d=4$, 
\bea
R(0)&\equiv&\widehat{\Pi}_{\mathrm{{L}}}(0) -\widehat{\Pi}_{\mathrm{{F}}}(0)\nonumber\\
&=&\frac{g^2}{(4\pi)^2}\left.\bigg\{\frac{m^4}6\frac{\partial}{\partial q^2}B_0(q^2;m^2,0)+ 3\left[B_0(q^2;m^2,0)-B_0(q^2;m^2,m^2)\right]\bigg\}\right\vert_{q^2=0}.
%\frac{g^2}{(4\pi)^2}\frac{m_0^2}{2(d-1)}\bigg[m_0^2\left.\frac{\partial}{\partial q^2}B_0(q^2;m_0^2,0)\right\vert_{q^2=0}\nonumber \\
%&+& \left(8d-14\right)B_0(0;m_0^2,0)-\left(4d+2\right)B_0(0;m_0^2,m_0^2)\bigg]
\label{R01}
\eea

We next 
extend the one-loop expression given in (\ref{R01}) to the non-perturbative regime,  
by introducing the following approximations:   
(i) we replace the (tree-level)  massive propagators appearing in  the function $B_0$
by their  fully dressed counterpart $\Delta$ (in  Landau gauge), and
(ii) the ``hard'' mass $m$  by its running counterpart. Then we find 
(in Euclidean space)

\bea
\left.\frac{\partial}{\partial q^2}B_0(q^2;m^2,0)\right\vert_{q^2=0}&\to&-\left.\frac1{\pi^2}\frac{\partial}{\partial q^2 }\int_{k }\frac{\Delta (k^2 )}{(k +q )^2}\right\vert_{q^2 =0}= \ \frac12\Delta(0),\nonumber \\
\left.B_0(q^2;m^2,0)-B_0(q^2;m^2,m^2)\right\vert_{q^2=0}&\to&\frac{1}{\pi^2}\int_{k }\frac{m^2(k^2 )\Delta^2(k ^2)}{k ^2}.
\eea
Thus, in the $SU(3)$ case one obtains the final result (with $y=k^2$)
\be
R(0)=
\frac32\ \frac{\alpha(\mu^2)m_0^2}{4\pi} \left[\frac{m_0^2}{12}\Delta(0)+3\int\! dy\, m^2(y)\Delta^2(y)\right]\,,
\label{R0}
\ee
where the multiplicative factor of $3/2$ corresponds 
to the ratio of the Casimir eigenvalues for the adjoint representations of the 
gauge groups $SU(3)$ and $SU(2)$.  

Since in Euclidean space $R(q^2)$ changes sign, we obtain\footnote{To go to Euclidean space, 
we set  $q^2 = -q^2_{\chic{\mathrm{E}}}$, with  $q^2_{\chic{\mathrm{E}}} >0$ 
the positive square of a 
Euclidean four-vector, define the Euclidean propagator as $\Delta_{\chic{\mathrm{E}}} (q^2_{\chic{\mathrm{E}}}) = -\Delta (-q^2_{\chic{\mathrm{E}}})$, 
and the integration measure as $\int_k = i \int_{k_{\chic{\mathrm{E}}}}$.  To avoid notational clutter we always suppress the subscript ``E''.}
\bea
\widehat{\Delta}_{\mathrm{{F}}}(q^2 )
&=&\frac1{q^2 +\widehat{\Pi}_{\mathrm{{F}}} (q^2 )}
\ = \
\frac1{(q^2 +\widehat{\Pi}_{\mathrm{{L}}}(q^2 ))
\left(1+\frac{R (q^2)}{q^2 +\widehat{\Pi}_{\mathrm{{L}}}(q^2 )}\right)}\nonumber \\
& = &
\frac{\widehat{\Delta}(q^2 )}{1+R (q ^2) \widehat{\Delta}(q^2 )},
\eea
arriving at the following relation  for the two couplings,
\be
\aptz=\frac{\alpha(0)}{1+ R (0) \widehat{\Delta}(0)}.
\label{map}
\ee

In order to get an approximate estimate for  
$\aptz$ we need to determine the value of $R (0)$ from (\ref{R0}). To that end, we use 
the lattice data for the $\Delta(y)$ appearing on the rhs, and a mass  $m^2(y)$  
that displays power-law running, given by (\ref{plr}). 
The   results   of    this   procedure   are   summarized   in
Table~\ref{tab1};  clearly, the values obtained for
$\aptz$  are  indeed  much  closer  to the expectations based on 
phenomenological studies.

We emphasize that 
Eq.~(\ref{R0})  constitutes  only a simplified estimate of the  complete
answer, and our results are suggestive at best. 
Note in particular that, as is evident from Table~\ref{tab1},  
Eq.~(\ref{map}) leads to the  introduction of  a spurious
dependence  on  the renormalization  scale  $\mu$  for the  ostensibly
RG-invariant quantity $\alpha_{\chic{\mathrm{PT}}}(q^2)$.

\begin{table}
	\centering
		\begin{tabular}{cccccc||c}
%\hline
%\hline
 $\;\Delta(0)\, [\mbox{GeV}^{-2}]\;$ & $\;\mu \,[\mbox{GeV}]\;$ & $\;\alpha(\mu^2)\;$ & $\;m_0 \,[\mbox{MeV}]\;$ &\;$\alpha(0)\;$ & $\;R(0)\;$& $\;\aptz\;$ \\
\hline
\hline
$5.51$ & $2.5$ & $0.467$ &$600$ &$6.40$  &$0.31$& $0.47$\\
$7.00$ & $4.0$ & $0.309$ &$600$ &$6.40$&$0.33$& $0.30$\\
$5.51$ & $2.5$ & $0.467$ &$500$ &$4.45$  &$0.15$& $0.62$\\
$7.00$ & $4.0$ & $0.309$ &$500$ &$4.45$ &$0.16$& $0.40$\\
\end{tabular}
\caption{\label{tab1}The gauge-invariant and universal IR fixed point $\aptz$ obtained from the Landau gauge $\alpha(0)$ one after applying  Eq.~(\ref{map}).}
\end{table}

\section{Conclusions and outlook}

In this article we have shown how 
to extract effective QCD charges from the available (quenched) lattice data for 
some of the fundamental Green's functions of QCD.  
We use two different definitions of the effective charge, 
whose construction follows a similar procedure, relying on the 
construction of RG-invariant quantities out of the 
judicious combination of the various field-theoretic ingredients.  
The effective charges obtained display the characteristic feature of
freezing  at a common finite (non-vanishing) value in the deep 
IR, as expected from a variety of theoretical 
and phenomenological considerations.

In addition, we have offered a plausible explanation 
for the observed discrepancy in the freezing values 
of the effective charges obtained from the lattice and those 
derived from the fitting of various QCD processes, sensitive to 
non-perturbative physics.  
Our claim is that the underlying reason for the discrepancy is the 
difference in the gauges (Landau vs Feynman) used in the two approaches.
We have studied this issue in the context of a toy model, which seems 
to corroborate this assertion.

It is clearly highly desirable 
to have available lattice  results for
the  gluon and  ghost  propagators  in gauges  other  than the  Landau.
In fact, a new gauge-fixing algorithm that may allow one to carry out 
lattice simulations in general  $R_\xi$ gauges 
has been recently proposed~\cite{Cucchieri:2009kk}.  
In addition, it is of considerable theoretical importance to 
obtain lattice results 
in the Feynman gauge of the BFM~\cite{Dashen:1980vm}, 
where, by  virtue of the  PT, quantities such as  the gluon propagator 
acquire a gauge-invariant  and universal status.  
Lattice results in this class of gauges would allow not only a direct
determination of  the phenomenologically relevant  coupling $\aptz$, but
will furnish a stringent test 
of the  SDE predictions    for     the
gluon~\cite{Aguilar:2009pp} and ghost propagators~\cite{Aguilar:2007nf}.

\section*{Acknowledgments}
 
We would like to thank  A. Cucchieri, T. Mendes, M. M\"uller-Preussker, 
and  O. Oliveira for kindly making their lattice
results available to us, and for their useful comments.  
The research of J.~P. is supported by the European FEDER and  Spanish MICINN under grant FPA2008-02878, and the Fundaci\'on General of the UV. The work of  A.C.A  is supported by the Brazilian Funding Agency CNPq under the grant 305850/2009-1.

\end{document}